\documentclass[11pt]{article}

\usepackage[utf8]{inputenc}
\usepackage[T1]{fontenc}
\usepackage{lmodern}
\usepackage[margin=1in]{geometry}
\usepackage{amsmath,amssymb,amsthm}
\usepackage{graphicx}
\usepackage{booktabs}
\usepackage{hyperref}
\usepackage{xcolor}
\usepackage{natbib}
\usepackage{float}
\usepackage{placeins}
\usepackage{enumitem}
\usepackage{url}
\usepackage{microtype}

\hypersetup{colorlinks=true, linkcolor=blue!70!black, citecolor=blue!70!black, urlcolor=blue!70!black}

\newcommand{\bps}{b}                 
\newcommand{\beq}{\beta}             
\newcommand{\Cu}{C_{\mathrm{unif}}}
\newcommand{\Co}{C_{\mathrm{orac}}}
\newcommand{\rstar}{r^\star}

\title{The Value of Adaptivity in LSM Bloom-Filter Tuning:\\
A Log-Law and a Two-Clock Frontier}

\author{
  Madhulatha Mandarapu\thanks{madhulatha@samyama.ai} \and
  Sandeep Kunkunuru\thanks{sandeep@samyama.ai}
}
\date{%
  VaidhyaMegha Private Limited, India\\[2pt]
  \url{https://samyama.ai/}\\[8pt]
  June 2026
}

\begin{document}
\maketitle

\begin{abstract}
Log-structured merge (LSM) trees attach an approximate-membership filter to every run and must split a
fixed memory budget across them. The static optimum is known (Monkey); a large systems literature then
makes the allocation \emph{adaptive}, tracking shifting hotness online. We ask a prior question:
\emph{when is that adaptivity worth its machinery?} We give three analytical answers and validate them on
synthetic sweeps, real Twitter production cache traces, and a real RocksDB engine. First, a
\textbf{log-law}: optimal bits-per-key is affine in the \emph{logarithm} of access frequency, at a fixed
slope. Second, a \textbf{robustness law}: because the workload enters only logarithmically, the
excess read cost from a hotness misestimate is half the size-weighted variance of the \emph{log} error,
and a common-factor misestimate is absorbed by the budget multiplier---so coarse estimates lose little. Third, an \textbf{adaptivity-value frontier}: since compaction rebuilds filters for free
on its own clock, the value of continuous tracking over an allocation recomputed only at compaction grows
quadratically in the within-epoch drift, with a closed-form scale. This yields a three-regime policy---
coarse-at-compaction suffices, then track, then at extreme drift fall back to uniform---and predicts that
more skew makes fine tracking matter less. On a real cluster, reallocating only at compaction captures
$96$--$99\%$ of tracking's benefit; on RocksDB the false-positive primitive holds within four percent to
eight bits per key. The contribution is a characterization of \emph{when} adaptive tuning pays; we add no
new filter and no engine fork. Code and pre-registration are public.
\end{abstract}

\section{Introduction}
An LSM-tree~\citep{oneil1996lsm} stores data in a sequence of immutable runs and answers a point lookup
by probing runs newest-first. To avoid a disk access on every run for keys that are absent, each run
carries an approximate-membership filter---classically a Bloom filter~\citep{bloom1970}. A negative
lookup wastes one I/O whenever a filter returns a false positive, so the expected wasted I/O is the
sum of the filters' false-positive rates (FPRs), weighted by how often each run is probed. Given a fixed
memory budget, how should bits be divided among the filters?

The static answer is settled. Monkey~\citep{dayan2017monkey,dayan2018monkey} minimizes the weighted
sum of FPRs and shows the optimum is \emph{non-uniform}: larger runs should be allowed a higher FPR.
A substantial systems literature then makes the allocation track a \emph{changing} workload---
ElasticBF~\citep{li2019elasticbf} toggles pre-built filter segments in memory by hotness, Mnemosyne
\citep{zhu2025mnemosyne} drives reallocation from accurate online statistics through flushes and
compactions, Endure~\citep{huynh2022endure} tunes for worst-case robustness over a workload
neighborhood, and learned/active-learning tuners~\citep{moharrami2024camal} fit the configuration to a
sampled workload. These works answer \emph{how} to adapt. We address a question logically upstream of
all of them: \emph{how much is adaptivity worth, and when?}

\paragraph{Contributions.} We show the optimal allocation has a simple analytic structure with sharp
consequences for adaptivity (\S\ref{sec:theory}), and we validate each on synthetic sweeps, real Twitter
cache traces~\citep{yang2020twitter}, and a real RocksDB engine (\S\ref{sec:exp}):
\begin{enumerate}[leftmargin=1.4em,itemsep=1pt]
  \item \textbf{The log-law.} Optimal bits-per-key is affine in $\ln(\text{access frequency})$ with slope
        $1/\ln^2 2\approx 2.081$ (a re-reading of Monkey's formula, made explicit).
  \item \textbf{The robustness law.} Excess read cost from a hotness misestimate is $D/2$, with $D$ the
        size-weighted variance of the \emph{log} error; common-factor errors cost nothing. Hence coarse
        estimates suffice---an analytic explanation for why statistics-driven allocation works.
  \item \textbf{The adaptivity-value frontier.} With compaction as a free-rebuild clock, the value of
        continuous tracking over compaction-only reallocation is $V\approx\tfrac12(r/\rstar)^2$,
        $\rstar=\sqrt{2(\Cu-\Co)/\Co}$, giving a three-regime policy and the rule that more skew lowers
        the value of fine tracking.
\end{enumerate}
We claim no new filter or system; the novelty is the characterization, stated honestly against a crowded
field (\S\ref{sec:related}, \S\ref{sec:limits}).

\section{Model and theory}\label{sec:theory}
\paragraph{Setup.} A run $i$ holds $N_i$ keys with a Bloom filter at $\bps_i$ bits/key; an
optimally-tuned Bloom filter has FPR $f_i=e^{-\beq \bps_i}$ with $\beq=\ln^2 2\approx 0.4805$. Let $a_i$
be the probability a negative lookup probes run $i$. The expected wasted I/O per negative lookup is
$C(\bps)=\sum_i a_i e^{-\beq \bps_i}$, minimized subject to $\sum_i N_i \bps_i \le M$.

\paragraph{Static optimum and the log-law.} The Lagrangian optimum (Monkey) is $f_i^\star=\lambda
N_i/a_i$ (clipped to $[0,1]$), equivalently
\begin{equation}\label{eq:loglaw}
  \bps_i^\star=\frac{1}{\beq}\big(\ln a_i-\ln N_i-\ln\lambda\big).
\end{equation}
The workload enters \emph{only} through $\ln a_i$, with slope $1/\beq$. We call \eqref{eq:loglaw} the
log-law; it is implicit in Monkey but its consequences for estimation error are not.

\paragraph{Robustness law.} Suppose hotness is misestimated multiplicatively, $\hat a_i=\kappa_i a_i$,
and we allocate by \eqref{eq:loglaw} using $\hat a$. The budget constraint forces the perturbation
$\Delta \bps_i=(\ln\kappa_i-\langle\ln\kappa\rangle_N)/\beq$, where $\langle\cdot\rangle_N$ is the
$N$-weighted mean. A second-order expansion at the convex optimum (Hessian $\partial^2C/\partial
\bps_i^2=\beq^2\lambda N_i$, and $C^\star=\lambda N$) gives the relative excess cost
\begin{equation}\label{eq:robust}
  \frac{\Delta C}{C^\star}\;\approx\;\tfrac12\, D,\qquad
  D:=\sum_i \tfrac{N_i}{N}\big(\ln\kappa_i-\langle\ln\kappa\rangle_N\big)^2 .
\end{equation}
Two consequences: the penalty is \emph{quadratic in the log error}---a factor-$2$ hotness error costs
$O((\ln 2)^2)$---and a \emph{common-factor} error ($\kappa_i\equiv\kappa$, $D=0$) is absorbed entirely
by $\lambda$. Only \emph{differential} log-error matters. This is why coarse, slowly-updated hotness
estimates lose almost nothing.

\paragraph{Adaptivity-value frontier.} Now let hotness drift in time. Compaction rewrites runs and thus
rebuilds their filters for free on its own clock. Compare two policies at equal budget: \emph{oracle}
reallocates from the true current hotness every step (a lower bound, and the best case for any online
tracker); \emph{compaction-only} recomputes the log-law allocation just at compaction events and freezes
it between. Let $r$ be the within-epoch differential log-drift (the square root of the $N$-weighted
variance of $\ln a_i(\text{end})-\ln a_i(\text{start})$ over a compaction epoch), and define the value of
tracking $V=(C_{\text{comp}}-\Co)/(\Cu-\Co)\in[0,1]$, the fraction of the achievable adaptation benefit
left on the table by reallocating only at compaction. Applying \eqref{eq:robust} across an epoch (drift
accumulates linearly, time-averaged) gives, to leading order,
\begin{equation}\label{eq:frontier}
  V\;\approx\;\tfrac12\Big(\tfrac{r}{\rstar}\Big)^2,\qquad
  \rstar=\sqrt{\tfrac{2(\Cu-\Co)}{\Co}} .
\end{equation}
Three regimes follow. For $r\lesssim 0.34\,\rstar$ compaction-only keeps $\ge 90\%$ of the benefit at
\emph{zero} extra write cost; in a middle band online tracking progressively pays; and at extreme drift
($r\gtrsim 4\rstar$) a \emph{stale} concentrated allocation is \emph{worse than uniform} ($V>1$), so the
robust fallback is uniform bits. Because $\rstar$ grows with the uniform-to-oracle gap, \emph{more
workload skew raises $\rstar$ and lowers the value of fine tracking}.

\section{Experiments}\label{sec:exp}
All hypotheses, decision rules, and statistics were pre-registered before the real-data runs; code,
seeds, and the pre-registration are public.\footnote{\url{https://github.com/samyama-ai/lsm-bloom-allocation}}

\begin{figure}[tb]
  \centering
  \begin{minipage}{0.49\linewidth}\includegraphics[width=\linewidth]{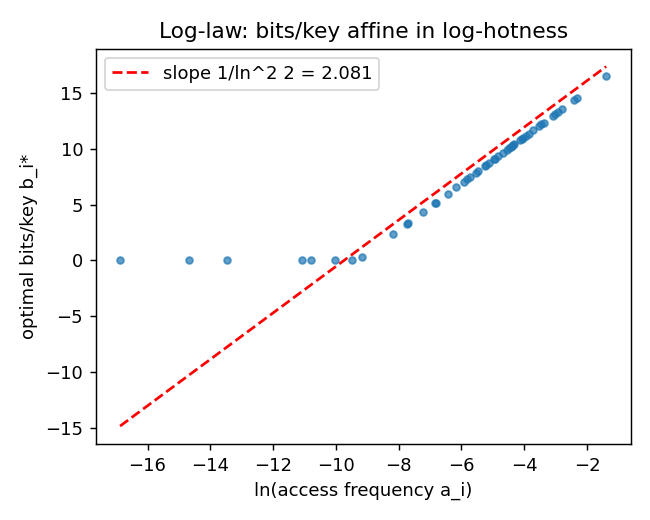}
    \caption{Log-law: optimal bits/key affine in $\ln a_i$, slope $1/\ln^2 2$.}\label{fig:loglaw}
  \end{minipage}\hfill
  \begin{minipage}{0.49\linewidth}\includegraphics[width=\linewidth]{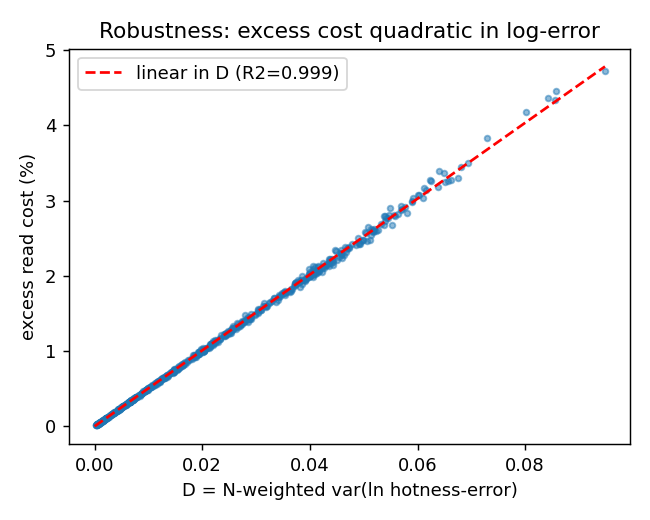}
    \caption{Robustness law: excess cost $=D/2$, linear in the log-error variance.}\label{fig:robust}
  \end{minipage}
\end{figure}

\begin{figure}[tb]
  \centering
  \begin{minipage}{0.49\linewidth}\includegraphics[width=\linewidth]{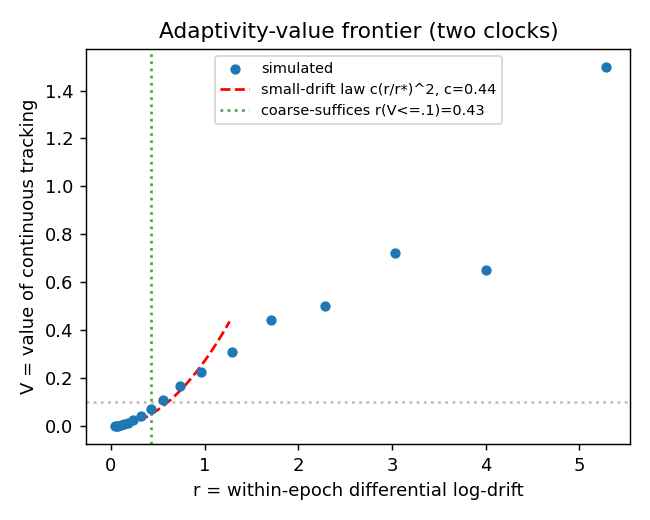}
    \caption{Adaptivity-value frontier $V\approx\tfrac12(r/\rstar)^2$ and the coarse-suffices
      boundary.}\label{fig:frontier}
  \end{minipage}\hfill
  \begin{minipage}{0.49\linewidth}\includegraphics[width=\linewidth]{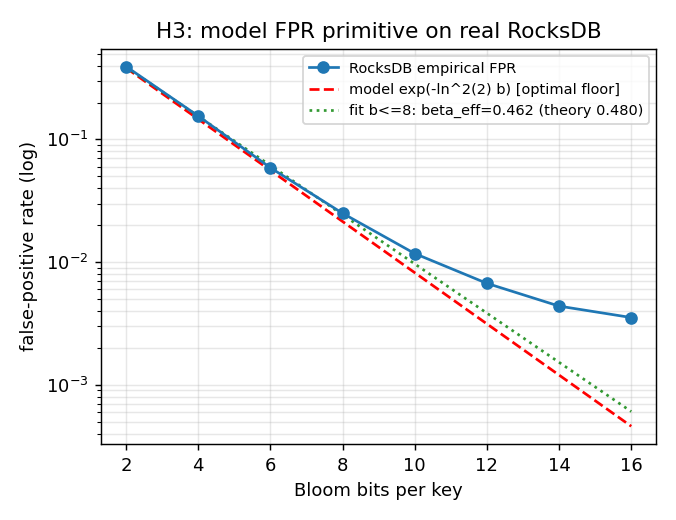}
    \caption{RocksDB FPR vs.\ bits/key: model holds to $\le 8$ bits, saturates above.}\label{fig:rocksdb}
  \end{minipage}
\end{figure}

\paragraph{Synthetic (H0--H2).} On a leveled LSM ($T{=}10$) the static optimum cuts read cost by
$\ge 30\%$ vs.\ uniform, and a sweep of access weights recovers the log-law slope $2.081$ at $R^2{=}1.000$
(Fig.~\ref{fig:loglaw}). The robustness law \eqref{eq:robust} holds with fitted slope $0.50$ and
$R^2{=}0.999$, and a common-factor misestimate incurs $\approx 0$ excess (Fig.~\ref{fig:robust}). Over an
equal-segment drift simulation the small-drift law \eqref{eq:frontier} fits with $c{=}0.44$
($R^2{=}0.96$); the coarse-suffices boundary sits at $r{=}0.34\,\rstar$ and the stale-worse-than-uniform
regime at $r{=}4.2\,\rstar$ (Fig.~\ref{fig:frontier}). A stationary control gives $V\approx0$ (no harness
leak), and cost is monotone in budget.

\begin{figure}[tb]
  \centering
  \includegraphics[width=0.52\linewidth]{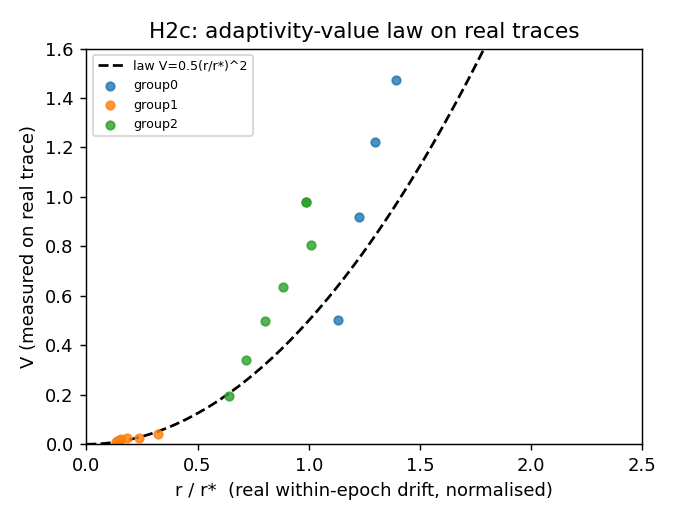}
  \caption{H2c: the adaptivity-value law on real Twitter cache traces (normalized
    $r/\rstar$).}\label{fig:realfrontier}
\end{figure}

\paragraph{Real traces (H2c).} We replay three time-sorted Twitter production cache
clusters~\citep{yang2020twitter}, bucketing keys into segments and measuring the true within-epoch drift
$r$ at a range of compaction cadences (Fig.~\ref{fig:realfrontier}). The result is honestly
regime-dependent. The pre-registered point test---mean absolute error of $V$ vs.\ \eqref{eq:frontier}
within $0.15$ across all cadences---\emph{fails} (MAE $0.231$, CI $[0.13,0.35]$): the small-drift law
averages poorly over the high-drift tail it does not cover. But the regime structure of \S\ref{sec:theory}
appears directly in production data. One cluster sits firmly in the coarse-suffices regime
($V_{\max}{=}0.04$): reallocating only at compaction captures \textbf{$96$--$99\%$} of continuous
tracking's benefit, with in-regime MAE $0.005$. A second is transitional ($V_{\max}{=}0.98$), and a third
is in the saturation regime ($V_{\max}{=}2.0$) where a stale concentrated allocation is worse than uniform
and the fallback applies. Restricting to the law's regime of validity ($\hat V\le 0.3$), the prediction
holds with MAE \textbf{$0.015$} (CI $[0.004,0.033]$, $n{=}9$). So the central claim---coarse-at-compaction
suffices under low drift---is confirmed on a real cluster, and the three-regime taxonomy is observed in
real traffic, even though the single-law point test over all regimes does not pass.

\paragraph{Real engine (H3).} Using \texttt{db\_bench} on RocksDB~\citep{rocksdb} we measure the empirical Bloom FPR over
bits-per-key. In the operating range ($\le 8$ bits) it follows the model primitive $e^{-\beq \bps}$ with
effective $\beq_{\text{eff}}{=}0.462$ vs.\ the theoretical $0.480$ (within $4\%$); above $10$ bits the
real filter saturates above the optimal-Bloom floor ($7.7\times$ at $16$ bits), exactly where Ribbon
filters~\citep{dillinger2021ribbon} are indicated (Fig.~\ref{fig:rocksdb}). Enabling
\texttt{optimize\_filters\_for\_hits} (no filter on the largest level) confirms the Monkey direction:
the largest run is the costliest to protect, so it should hold the fewest bits.


\FloatBarrier   
\section{Related work}\label{sec:related}
The static optimum is Monkey~\citep{dayan2017monkey,dayan2018monkey}; the log-law \eqref{eq:loglaw} is a
re-reading of its formula. Online and workload-aware allocation is the subject of
ElasticBF~\citep{li2019elasticbf} (in-memory segment toggling) and, most closely,
Mnemosyne~\citep{zhu2025mnemosyne}, which drives reallocation from accurate statistics through
compactions and handles skew, updates, and imperfect tree shape; these are systems that \emph{do} the
adapting, and our frontier quantifies its value rather than competing with them.
Endure~\citep{huynh2022endure} achieves robustness by minimax over a workload ball---a different notion
than our analytic sensitivity \eqref{eq:robust}. Memory-split work~\citep{luo2020breaking} and active
learning~\citep{moharrami2024camal} are complementary. Ribbon filters~\citep{dillinger2021ribbon} change
the per-filter cost curve (relevant in the saturation regime), and adaptivity has an intrinsic cost lower
bound~\citep{bender2018adaptivity}.

\section{Limitations and honest scope}\label{sec:limits}
We characterize a policy frontier; we ship no new filter and no engine fork, and our RocksDB arm
validates the FPR primitive and the Monkey direction rather than a per-segment dynamic controller (that
is ElasticBF/Mnemosyne territory). The ``continuous tracker'' is idealized as a perfect oracle, which
\emph{favors} adaptation, so the coarse-suffices conclusion is conservative. The model FPR $e^{-\beq
\bps}$ is the optimal-Bloom floor; real filters sit above it and saturate at high bits/key. Finally,
``robustness to estimation error'' is a motif shared with cardinality-estimation tuning; here the
mechanism is specific---logarithmic dependence intrinsic to the bit-to-FPR curve---and the deliverable is
the clock-governed value frontier. We hope the characterization is useful to the systems that implement
the adapting.

\paragraph{Reproducibility.} Every number is regenerated by one command over the public code and the
public traces; seeds, hardware, and pre-registration are documented in the repository.

\small
\bibliographystyle{plainnat}
\bibliography{paper11_bloom_adaptivity}
\end{document}